\pdfoutput=1
\documentclass[11pt,a4paper]{article}
\usepackage{amsmath,amssymb,url,hyperref,graphicx,cite}

\begin{document}

\begin{center}
{\large \bf 
Lepton Electric Dipole Moment and Strong CP Violation
}

\vskip 1.2cm
Diptimoy Ghosh$^{1,2}$ and Ryosuke Sato$^2$
\vskip 0.4cm

$^1${\it
International Centre for Theoretical Physics, Strada Costiera 11, 34014 Trieste, Italy
}
$^2${\it
Department of Particle Physics and Astrophysics,\\ Weizmann Institute of Science, Rehovot 7610001, Israel
}

\vskip 1.5cm

\abstract{
Contribution of the strong CP angle, $\bar\theta$, to the Wilson Coefficients of electron and muon electric dipole 
moment (EDM) operators are discussed.
Previously, $\bar\theta$ contribution to the electron EDM operator was calculated by Choi and Hong \cite{Choi:1990cn}.  However, the effect of CP-violating 
three meson coupling was missing in \cite{Choi:1990cn}.
We include this missing contribution for the first time in the literature, and reevaluate the Wilson coefficients of the lepton EDM operator.
We obtain 
$d_e = - (2.2-8.6) \times 10^{-28} \, \bar\theta$ e-cm which is 15 -- 70 \% of the result obtained in 
\cite{Choi:1990cn}. 
We also estimated the muon EDM as $d_\mu = - (0.5-1.8) \times 10^{-25} \, \bar\theta$ e-cm.
Using $|\bar\theta| \lesssim 10^{-10}$ suggested by the neutron EDM measurements, we obtain $|d_e| \lesssim 8.6 \times 
10^{-38}$ e-cm and $|d_\mu| \lesssim 1.8 \times 10^{-35}$ e-cm.
The $\bar\theta$ contribution to the muon EDM is much below the sensitivities of the current and near future experiments.
Our result shows that the $\bar\theta$ contribution to $d_{e,\mu}$ can be larger than the CKM contributions by many orders of magnitude.
}
\end{center}

\section{Introduction}
Precise measurement of EDMs is an important probe of CP violation. In particular, the lepton EDMs have recently 
received much attention because the experimental sensitivity is expected to improve considerably in near future.
The electron EDM will be searched by ACME-II and III experiments \cite{ACMETalk}, and the muon EDM will be 
searched in J-PARC \cite{Iinuma:2011zz} and Fermilab \cite{Chislett:2016jau}.
These experiments will probe interesting regions of parameter space of many well motivated models beyond the 
Standard Model (SM) \cite{Abe:2014gua,Berger:2015eba,ACMETalk}. 

Before discussing EDMs in models beyond the SM,  it is important to first know the SM predictions precisely.
The SM has two possible sources of CP violation: the phase of Cabibbo-Kobayashi-Maskawa (CKM) matrix and 
the QCD $\theta$ term (by QCD $\theta$ term, we always refer to the field redefinition independent 
combination, $\bar\theta$ ). 
In this paper, we discuss the Wilson Coefficient of the lepton EDM operator:\footnote{The electron EDM experiments 
actually measure the EDM of an atom or a molecule. If we consider the effect of $\bar\theta$ to such measurements,
CP violating electron-nucleon interaction $(\bar e i\gamma^5 e)(\bar N N)$ gives the dominant contribution,
and the true lepton EDM operator contribution (i.e., from $d_e$) is negligible. See \textit{e.g.}, refs.~\cite{Dedes:1999uj,
Chupp:2014gka}.
On the other hand, the muon EDM experiments measure spin precession of a single muon. 
Hence, unlike the electron EDM experiments, these experiments actually measure the Wilson Coefficient 
$d_\mu$.
}
\begin{align}
{\cal L}_{\rm eff.}
\ni
-\frac{i}{2} d_\ell  \, \bar\ell \sigma_{\mu \nu} \gamma_5 \ell F^{\mu\nu}  \, .
\label{edm-def}
\end{align}
The CKM contribution to $d_e$ is $\simeq 10^{-44}$ e-cm, see for example, \cite{Pospelov:2013sca} and 
the references therein.
The $\bar\theta$ contribution to electron EDM was first discussed by Choi and Hong in \cite{Choi:1990cn}. They 
obtained  $d_e \simeq 1.4 \times 10^{-27} \bar\theta$ e-cm. In their work, they considered CP violating 
meson-baryon-baryon coupling in chiral Lagrangian \cite{Crewther:1979pi}, and estimated chiral logarithm 
contribution to $d_e$.
However, there also exists CP violating meson-meson-meson couplings, and contributions from these couplings 
were not discussed in \cite{Choi:1990cn}. 
In this paper, we calculate $d_e$ and $d_\mu$ taking into account both the CP violating meson-baryon-baryon and 
meson-meson-meson couplings.

In section \ref{sec:chpt}, we briefly review the chiral Lagrangian for the calculation of EDMs. 
The details of the calculations of $d_e$ and $d_\mu$ are presented in section \ref{sec:EDM}. 
In section \ref{sec:conclusion}, we briefly discuss our results and summarize.

\section{Chiral Lagrangian for the calculation of EDMs}
\label{sec:chpt}

In this section, we briefly review the  chiral Lagrangian for the calculation of EDMs induced by the the strong CP phase $\bar\theta$ 
\cite{Crewther:1979pi, Shifman:1979if} (see also \cite{Pich:1991fq}). 
According to the CCWZ prescription \cite{Coleman:1969sm, Callan:1969sn}, one introduces the coset fields $\xi_{L,R}$ and the 
baryon field $B$ transforming under $SU(3)_L \times SU(3)_R$ chiral symmetry as
\begin{align}
\xi_L \to g_L \xi_L h^\dagger,\qquad
\xi_R \to g_R \xi_R h^\dagger,\qquad
B \to h B h^\dagger \, .
\end{align}
where $h$ is a compensator field, $h = h(g_L, g_R, \xi_L \xi_R^\dagger) \in SU(3)_V$ which satisfies 
$h(g,g,\xi_L \xi_R^\dagger) = g$.
The Meson field $U$ is defined by the coset fields as $U \equiv \xi_L \xi_R^\dagger$.
The quark mass matrix $M$ is introduced as a spurion for explicit $SU(3)_L \times SU(3)_R$ breaking.
The transformation rules for $U$ and $M$ are
\begin{align}
U \to g_L U g_R^\dagger \, ,\qquad
M \to g_L M g_R^\dagger \, .
\end{align}

\subsection{Meson}

The Meson field $U$ is written as 
\begin{align}
U = \exp\left( -\frac{\sqrt{2}i\phi}{f_\pi} \right),\qquad
\phi =
\left(\begin{array}{ccc}
\pi^0/\sqrt{2} + \eta^0/\sqrt{6} & \pi^+                             & K^+ \\
\pi^-                            & -\pi^0/\sqrt{2} + \eta^0/\sqrt{6} & K^0 \\
K^-                              & \bar K^0                          & -2\eta^0/\sqrt{6} \\
\end{array}\right),
\end{align}
and the $SU(3)_L \times SU(3)_R$ invariant Lagrangian for the Mesons is 
\begin{align}
{\cal L}_{\rm meson} = \frac{f_\pi^2}{4} {\rm tr}\left[ D_\mu U D^\mu U^\dagger \right]
                     +\frac{f_\pi^2}{2} B_0 {\rm tr}\left[ M^\dagger U + U^\dagger M \right], \label{eq:meson lagrangian}
\end{align}
where $f_\pi = 93~{\rm MeV}$. The constant $B_0$ is determined by the pion and quark masses.
We write the quark mass matrix $M$ as
\begin{align}
M = {\rm diag}(m_u, m_d, m_s) + i\bar\theta m_* I,\qquad
m_* = 1/(m_u^{-1} + m_d^{-1} + m_s^{-1}).
\end{align}
The imaginary part of $M$ has the role of a spurion for explicit $P$ and $CP$ breaking.
Since ${\rm Im}(M)$ in this basis does not break $SU(3)_V$ flavor symmetry, we do not have tadpole term 
for $\pi^0$ and $\eta^0$ \cite{Baluni:1978rf, Crewther:1979pi}.

\subsubsection*{Meson masses}

By expanding  the term $(f_\pi^2/2) B_0 {\rm tr}[M^\dagger U + U^\dagger M]$ in ${\cal L}_{\rm meson}$ we obtain the meson 
masses as 
\begin{align}
m_\pi^2 = B_0(m_u + m_d),\quad
m_{K^{\pm,0}}^2 = B_0(m_{u,d} + m_s),\quad
m_\eta^2 = \frac{1}{3}B_0(m_u + m_d + 4m_s).
\end{align}
Therefore, $B_0 = m_\pi^2/2\bar m$ where $\bar m = (m_u+m_d)/2$.

\subsubsection*{CP violating couplings of mesons}

The term ${\cal L}_{\rm meson} \ni (f_\pi^2/2) B_0 {\rm tr}[M^\dagger U + U^\dagger M]$ also gives CP violating interaction 
terms involving mesons. 
We obtain the triple pion coupling to be 
\begin{align}
{\cal L}_{3\pi}
~=~ \frac{\bar\theta B_0 m_*}{3f_\pi} \pi^a \pi^b \pi^c d_{abc}
~=~ \frac{\bar\theta m_*}{\bar m} \frac{m_\pi^2}{6f_\pi} \pi^a \pi^b \pi^c d_{abc} \, , 
\end{align}
where,  $d^{abc}$ is defined in terms of the Gell-Mann matrices $\lambda^a$ which satisfy $\{\lambda^a, \lambda^b\} = 
(4/3)\delta^{ab} + 2 d^{abc} \lambda^c$.  See also Refs.~\cite{Crewther:1979pi, Shifman:1979if} for an alternative derivation. 
For interactions involving the $\eta$ and $K$ mesons, we obtain
\begin{align}
{\cal L}_{3\pi}
=& \frac{\bar\theta m_*}{\bar m} \frac{m_\pi^2}{\sqrt{3} f_\pi}  \left( \eta \pi^+ \pi^- + \frac{1}{2}\eta \pi^0 \pi^0 \right)
   - \frac{\bar\theta m_*}{\bar m} \frac{m_\pi^2}{6\sqrt{3} f_\pi} \eta^3
   - \frac{\bar\theta m_*}{\bar m} \frac{m_\pi^2}{2\sqrt{3} f_\pi} \eta (K^+ K^- + K^0 \bar K^0 ) \nonumber\\
 & + \frac{\bar\theta m_*}{\bar m} \frac{m_\pi^2}{2f_\pi} \left( \pi^0  K^+ K^- - \pi^0 K^0 \bar K^0 + \sqrt{2}\pi^- K^+ \bar K^0 + \sqrt{2}\pi^+ K^- K^0 \right).
\end{align}
In section \ref{CPV-meson-gaga}, we will calculate the CP violating $\pi^0 F_{\mu\nu} F^{\mu\nu}$ and $\eta F_{\mu\nu} F^{\mu\nu}$ couplings
by using the above interaction terms.

\subsection{Baryon}

The Baryon octet field is decomposed as
\begin{align}
B=
\left(\begin{array}{ccc}
\Sigma^0/\sqrt{2} + \Lambda^0/\sqrt{6} & \Sigma^+                                & p \\
\Sigma^-                               & -\Sigma^0/\sqrt{2} + \Lambda^0/\sqrt{6} & n \\
\Xi^-                                  & \Xi^0                                   & -2\Lambda^0/\sqrt{6} \\
\end{array}\right) \, .
\end{align}
We define $\tilde M$ and $\tilde M_+$ as
\begin{align}
\tilde M = \xi_L^\dagger M \xi_R \,\, ,\qquad
\tilde M_+ = (\tilde M + \tilde M^\dagger )/2.
\end{align}
It can be easily seen that $\tilde M$ and $\tilde M_+$ transform as
\begin{align}
\tilde M \to h \tilde M h^\dagger,\qquad
\tilde M_+ \to h \tilde M_+ h^\dagger \, .
\end{align}

Since $\tilde M$ is transformed as $\tilde M\to \tilde M^\dagger$ under parity transformation, 
$\tilde M_+$ is parity invariant. 

\subsubsection*{Baryon masses}

It is always possible to choose a basis such that $\xi_L = \xi_R^\dagger$,
\textit{i.e.}, $\xi_L = U^{1/2}$ and $\xi_R = U^{-1/2}$.
In this basis,
\begin{align}
\tilde M_+ = \frac{1}{2}\left( U^{-1/2} M U^{-1/2} + U^{1/2} M^\dagger U^{1/2}  \right).
\end{align}
Baryon mass splitting is generated by the following parity invariant terms, 
\begin{align}
{\cal L}_{\rm baryon} \ni
-b_1 {\rm tr}[\bar B \tilde M_+ B]
-b_2 {\rm tr}[\bar B B \tilde M_+] \, . \label{eq:baryon mass term}
\end{align}
The mass splittings can now be written as
\begin{align}
m_\Lambda - m_N &= \left( \frac{2}{3}b_1 - \frac{1}{3} b_2 \right) (m_s-\bar m) \, ,\\
m_\Sigma - m_N &= -b_2 (m_s-\bar m)\, ,\\
m_\Xi - m_N &= (b_1-b_2) (m_s-\bar m)\, .
\end{align}

Here we have defined $m_N = (m_p + m_n)/2$,
$m_\Sigma \equiv m_{\Sigma^0} = (m_{\Sigma^+} + m_{\Sigma^-})/2$,
and $m_\Xi = (m_{\Xi^0} + m_{\Xi^-})/2$ \, .
The parameters $b_1$ and $b_2$ can now be determined in terms of the Baryon masses. The are given by  
\begin{align}
b_1  = \frac{m_\pi^2}{2\bar m} \frac{m_\Xi - m_\Sigma}{m_K^2 - m_\pi^2} ,\qquad
b_2  = -\frac{m_\pi^2}{2\bar m} \frac{m_\Sigma - m_N}{m_K^2 - m_\pi^2}  \, . 
\end{align}

\subsubsection*{CP violating couplings of baryons}

The Lagrangian in Eq.~(\ref{eq:baryon mass term}) also gives rise to CP violating interaction terms between 
mesons and baryons. 
For example, CP violating interaction terms which are relevant to neutron EDM are
\begin{align}
{\cal L}_{\rm baryon} ~\ni~
	\sqrt{2} \bar\theta \tilde b_1 ( \pi^- \bar n p + \pi^+ \bar p n )
	+ \sqrt{2} \bar\theta \tilde b_2 ( K^+ \bar n \Sigma_- + K^- \bar \Sigma_- n ) \,  , 
\end{align}
where $\tilde b_1$ and $\tilde b_2$ are defined as
\begin{align}
\tilde b_1 &~=~ \frac{b_1 m_*}{f_\pi} ~=~ \frac{m_\pi^2}{2f_\pi}\frac{m_\Xi-m_\Sigma}{m_K^2-m_\pi^2} \frac{m_*}{\bar m} ~\approx~ 0.025 \, ,\\
\tilde b_2 &~=~ \frac{b_2 m_*}{f_\pi} ~=~ -\frac{m_\pi^2}{2f_\pi}\frac{m_\Sigma-m_N}{m_K^2-m_\pi^2} \frac{m_*}{\bar m} ~\approx~ -0.050 \, .
\end{align}
Here we took $m_u = 2.15~{\rm MeV}$, and $m_d = 4.70~{\rm MeV}$ \cite{Olive:2016xmw}.
The above formulae of $\tilde b_1$ and $\tilde b_2$ hold at the leading order of the chiral perturbation theory.
For the discussion on the next leading order corrections, see Ref.~\cite{deVries:2015una}.
The interaction terms of $\pi^0/\eta^0$ with the charged baryons induce the CP violating $\pi^0 F_{\mu\nu} F^{\mu\nu}$ and 
$\eta F_{\mu\nu} F^{\mu\nu}$ couplings. 
The relevant interaction terms are given by 
\begin{align}
{\cal L}_{\rm baryon} ~\ni~
	& \bar\theta \tilde b_1 \pi^0 \bar p p
	+ \bar\theta \tilde b_2 \pi^0 \bar \Xi_- \Xi_-
	+ \bar\theta (\tilde b_1 - \tilde b_2) \pi^0 (\bar \Sigma_+ \Sigma_+ - \bar \Sigma_- \Sigma_-)  \nonumber\\
	& \hspace*{-5mm} + \bar\theta \frac{\tilde b_1 - 2\tilde b_2}{\sqrt 3} \eta \bar p p
	+ \bar\theta \frac{-2\tilde b_1 + \tilde b_2}{\sqrt 3} \eta \bar \Xi_- \Xi_-
	+ \bar\theta \frac{(\tilde b_1 + \tilde b_2)}{\sqrt 3} \eta (\bar \Sigma_+ \Sigma_+ + \bar \Sigma_- \Sigma_-) \, . 
\end{align}

\section{Lepton electric dipole moment}
\label{sec:EDM}

\begin{figure}
\centering
\includegraphics[width=0.5\hsize]{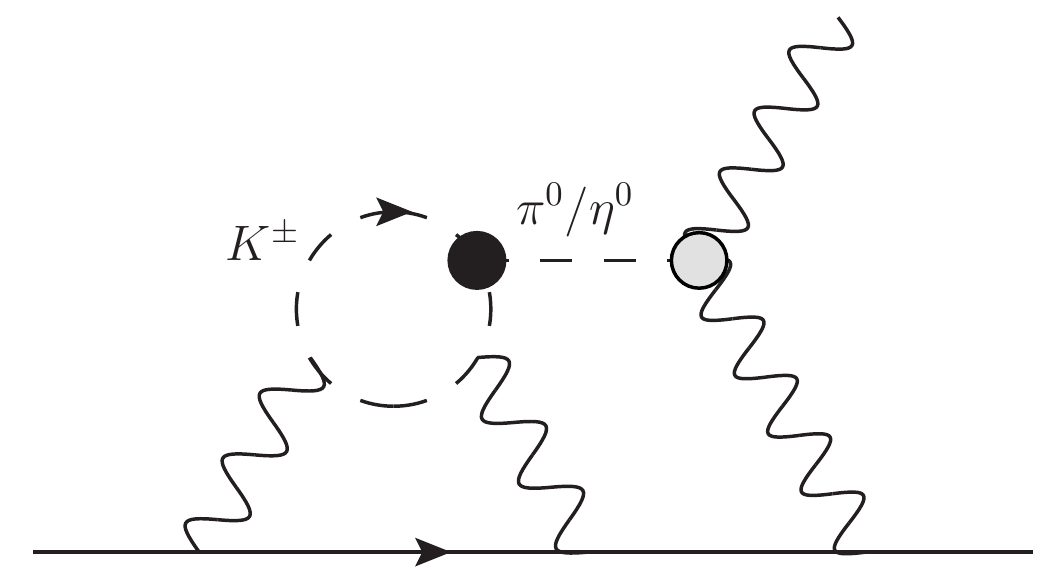}
\caption{
An example diagram that contributes to the lepton EDM.
The black circle is the CP-violating coupling and the grey circle is the CP-conserving interaction arising from the chiral anomaly matching.}
\label{fig:threeloop}
\end{figure}

In this section, we estimate the lepton EDM by using the chiral Lagrangian discussed in the previous section.
The dominant source of lepton EDM is hadronic light-by-light diagrams, such as the one shown in Fig.~\ref{fig:threeloop}.
In particular, we estimate the chiral logarithm contributions arising from this class of diagrams.

\subsection{CP violating meson-photon-photon couplings}
\label{CPV-meson-gaga}

As shown in Fig.~\ref{fig:threeloop}, the relevant diagrams have CP-violating ($\pi^0/\eta^0$)-photon-photon subdiagrams.
If the charged particles inside loop are heavy enough, these subdiagrams can be written as the following effective interactions, 
\begin{align}
{\cal L}_{\rm eff.} \ni
	\bar\theta c_\pi \frac{\alpha}{4\pi} \frac{\pi^0}{f_\pi} F_{\mu\nu} F^{\mu\nu}
	+ \bar\theta c_\eta \frac{\alpha}{4\pi} \frac{\eta^0}{f_\pi} F_{\mu\nu} F^{\mu\nu}.
\end{align}
The coefficients $c_\pi$ and $c_\eta$ are obtained by one loop calculation in the chiral Lagrangian.
For $c_\pi$, there are contributions from $K^\pm$ as well as the baryons ($p, \Xi^-$).
The $K^\pm$ contribution is denoted by $c_\pi^{(K)}$, and the baryon contribution is denoted by $c_\pi^{(b)}$.
We obtain
\begin{align}
c_\pi^{(K)} &= -\frac{1}{24} \frac{m_*}{\bar m}\frac{m_\pi^2}{m_K^2} ~\approx~ -1.41 \times 10^{-3}, \\
c_\pi^{(b)} &= 
		- \frac{2\tilde b_1}{3} \frac{f_\pi}{m_p}
		- \frac{2\tilde b_2}{3} \frac{f_\pi}{m_\Xi} ~\approx~ 0.74 \times 10^{-3},\\
c_\pi & = c_\pi^{(K)} + c_\pi^{(b)} \approx -0.68 \times 10^{-3} \, . 
\end{align}
The baryon-loop contribution $c_\pi^{(b)}$ was calculated in Ref.~\cite{Choi:1990cn},
and our result is consistent with the reference.
The sign of $c_\pi^{(K)}$ is opposite to $c_\pi^{(b)}$. 
So these contributions are destructive, and we will see in the next subsection that the resulting EDM is smaller than that of 
Ref.~\cite{Choi:1990cn}. 

For $c_\eta$, the one-loop baryon contribution is 
\begin{align}
c_\eta &= 
	- \frac{2( \tilde b_1-2\tilde b_2 )}{3\sqrt{3}} \frac{f_\pi}{m_p}
	- \frac{2( \tilde b_2-2\tilde b_1 )}{3\sqrt{3}} \frac{f_\pi}{m_\Xi}
	- \frac{4( \tilde b_1+\tilde b_2 ) }{3\sqrt{3}} \frac{f_\pi}{m_\Sigma} ~\approx~ -0.53 \times 10^{-3}.
\end{align}
Loop diagrams involving $\pi^\pm$ or $K^\pm$  also contribute to $\eta^0$-photon-photon amplitude however, 
it is not appropriate to consider their contribution as an effective $\eta^0 F_{\mu\nu} F^{\mu\nu}$ interaction because 
$\pi^\pm$ and $K^\pm$ are lighter than $\eta^0$.
We will comment on the $\pi^\pm$ and $K^\pm$ loop contributions in section \ref{eq:eta pi pi contribution}.

\subsection{Chiral logarithm for lepton EDM}

Here we discuss chiral logarithm contribution to lepton EDM.
As shown in Fig.~\ref{fig:threeloop}, the relevant diagrams have either $\pi^0$ or $\eta^0$ propagator.
Below, we discuss $\pi^0$ and $\eta^0$ exchange diagrams separately.

\subsubsection{$\boldsymbol\pi^0$ contribution}

\begin{figure}
\centering
\includegraphics[width=0.23\hsize]{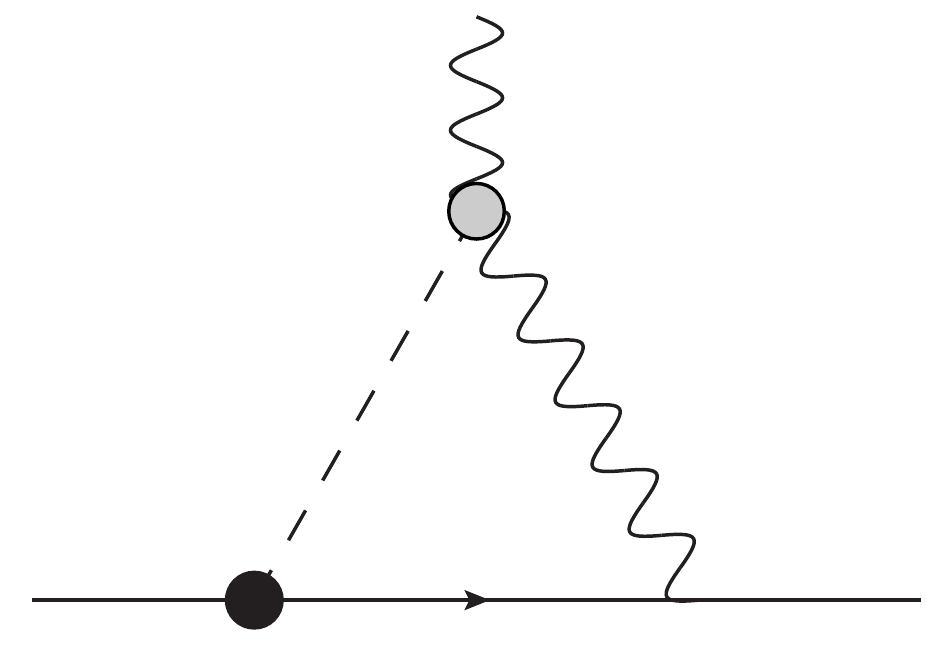}
\includegraphics[width=0.23\hsize]{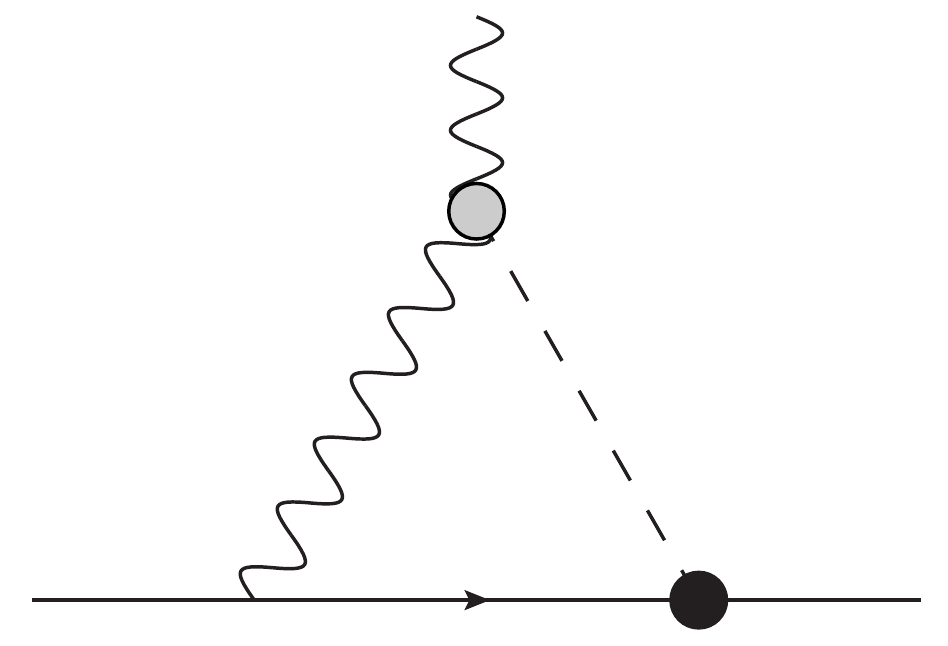}
\includegraphics[width=0.23\hsize]{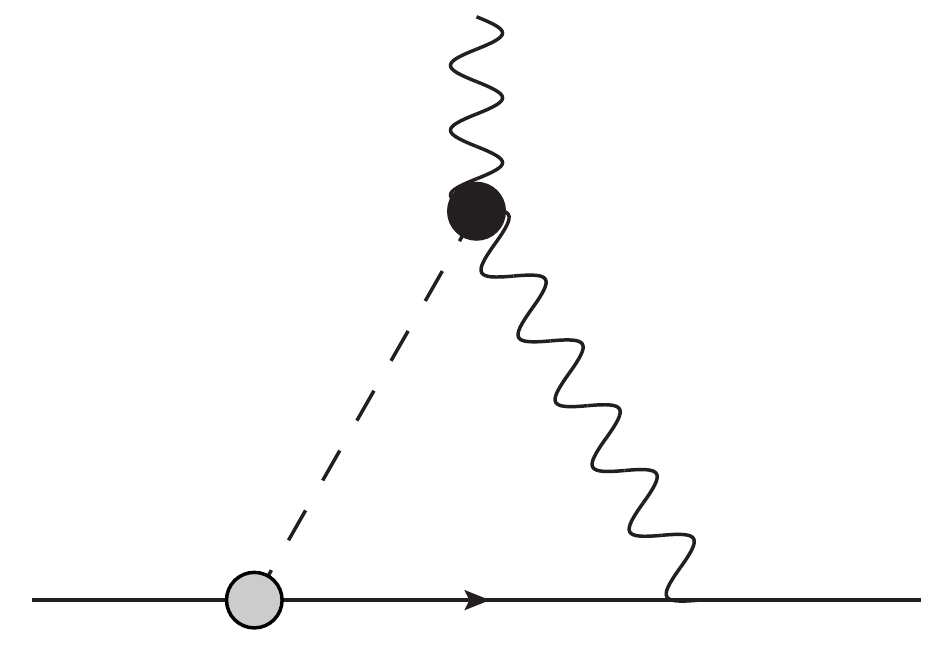}
\includegraphics[width=0.23\hsize]{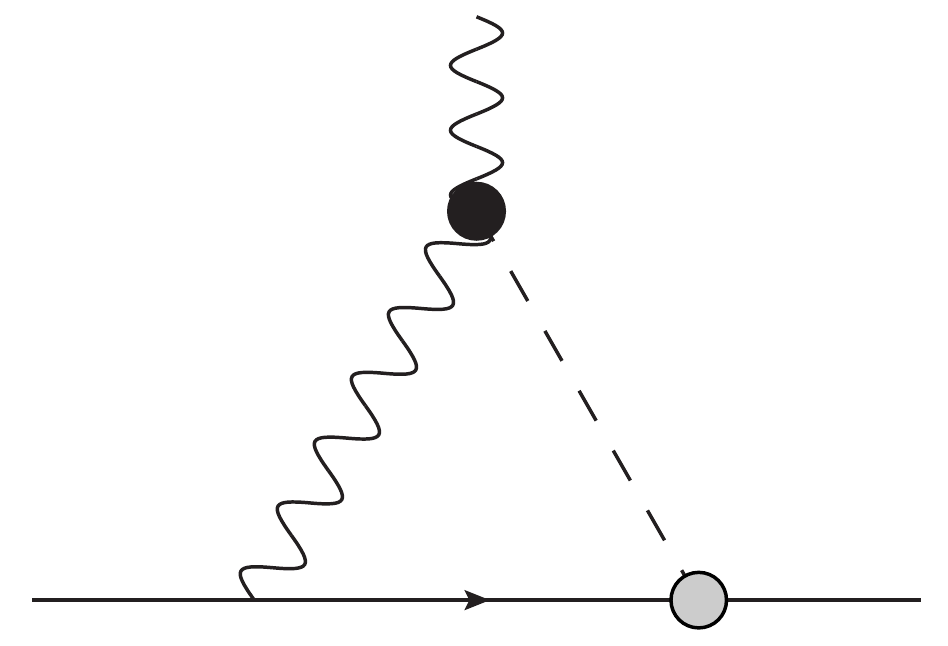}
\caption{Diagrams relevant for the RG equations in Eq.~(\ref{RG-eqns}). The black (grey) circle denotes the 
CP-breaking (CP-conserving) vertex. \label{RG-figs}}
\end{figure}

The effective interaction terms relevant for the calculation of $d_\ell$ are \cite{Choi:1990cn}
\begin{align}
{\cal L}_{\rm eff.}
\ni
-\frac{i}{2} d_\ell  \, \bar\ell \sigma_{\mu \nu} \gamma_5 \ell F^{\mu\nu}
+ A_1 \frac{\pi^0}{f_\pi} F_{\mu\nu} \tilde F^{\mu\nu}
+ A_2 \frac{\partial_\mu \pi^0}{f_\pi} \bar\ell \gamma^\mu \gamma^5 \ell 
+ C_1 \frac{\pi^0}{f_\pi} F_{\mu\nu} F^{\mu\nu} \nonumber \\
+ \, 2C_2 m_e \frac{\pi^0}{f_\pi} \bar\ell\ell \, ,
\end{align}
where $\sigma^{\mu\nu} = (i/2)[\gamma^\mu, \gamma^\nu]$.
The renormalization group (RG) equations for $d_\ell$, $A_2$, and $C_2$ are given by \cite{Choi:1990cn} 
(see Fig.~\ref{RG-figs} for the relevant diagrams)
\begin{align}
\frac{d d_\ell}{d\log\mu} = -\frac{e m_\ell}{\pi^2 f_\pi^2} (A_1 C_2 + A_2 C_1)\, ,\qquad
\frac{d A_2}{d\log\mu} = -\frac{3\alpha}{\pi}A_1 \, ,\qquad
\frac{d C_2}{d\log\mu} = -\frac{3\alpha}{\pi}C_1 \, . \label{RG-eqns}
\end{align}

The coefficient $A_1$ is determined by the chiral anomaly matching, and is given by 
\begin{align}
A_1 = \frac{\alpha}{4\pi} \, .
\end{align}
The coefficient $C_1$ was calculated in the previous section. They are given below for convenience, 
\begin{align}
C_1(\mu) = \bar\theta \frac{\alpha}{4\pi} \times 
\begin{cases}
c_\pi^{(b)}                & (\mu > m_K) \\
c_\pi^{(b)} + c_\pi^{(K)}  & (\mu < m_K) \, . \\
\end{cases}
\end{align}
As the kaons are integrated out at the scale of $m_K$,  $K^\pm$ loop contributions are included in $C_1(\mu)$ only for $\mu < m_K$.
Solving the RG equations with a boundary conditions, $A_2(\Lambda) = C_2(\Lambda)=0$, we obtain $A_2 (\mu)$ and $C_2 (\mu)$ as
\begin{align}
A_2 (\mu) &= 12 \left( \frac{\alpha}{4\pi} \right)^2 \log \frac{\Lambda}{\mu}\, ,\\
C_2 (\mu) &=
\begin{cases}
     12 \bar\theta c_\pi^{(b)} \left( \frac{\alpha}{4\pi} \right)^2 \log \frac{\Lambda}{\mu} & (\mu > m_K) \\
     12 \bar\theta c_\pi^{(b)} \left( \frac{\alpha}{4\pi} \right)^2 \log \frac{\Lambda}{\mu} 
        +12 \bar\theta c_\pi^{(K)} \left( \frac{\alpha}{4\pi} \right)^2 \log \frac{m_K}{\mu} & (\mu < m_K) \\
\end{cases} \, .
\end{align}
Using the above formulae, it is now straightforward to calculate the $\pi^0$ exchange contribution to the lepton EDM, $d_\ell^\pi$, 
which is given by 
\begin{align}
d_\ell^\pi(\mu)
= e \bar\theta \times \frac{12 m_\ell}{\pi^2 f_\pi^2}\left( \frac{\alpha}{4\pi} \right)^3
	\left[
		c_\pi^{(b)} \log^2 \frac{\Lambda}{\mu}
		+ c_\pi^{(K)} \log \frac{m_K}{\mu} \log \frac{\Lambda}{\mu}
	\right]. \label{eq:pion contribution}
\end{align}
The above contributions are the leading chiral logarithm terms. 
In order to calculate the next-to-leading terms, one has to determine the incalculable counter terms from some observables.
To avoid this procedure, here we estimate $d_e$ and $d_\mu$ by varying $\mu$ from $m_\pi/2$ to $m_\pi$ in 
Eq.~(\ref{eq:pion contribution}).
We take the cutoff scale $\Lambda$ as $4\pi f_\pi$.
We obtain
\begin{align}
d_e^\pi \simeq -(1.4-5.5) \times 10^{-28} \bar\theta \, \, \text{e-cm}, \qquad
d_\mu^\pi \simeq -(0.3-1.1) \times 10^{-25} \bar\theta \, \, \text{e-cm}. \label{eq:pion contribution(numerical)}
\end{align}
The contribution of $c_\pi^{(b)}$ in Eq.~(\ref{eq:pion contribution}) corresponds to the contribution which is calculated by Ref.~\cite{Choi:1990cn}.
The size of $d_e^\pi$ is smaller than that
because $K^\pm$ loop and the baryon loop contributions are destructive.

Before closing this section, let us comment on the $Z$ boson exchange diagrams. Although $Z$ boson does not contribute to 
the light-by-light diagram, $Z$ boson exchange gives a contribution to $\pi^0 e^+ e^-$ coupling at tree level, 
generating the coupling $A_2 = (1/\sqrt{2}) G_F f_\pi^2$ \cite{Choi:1990cn}.
The resulting contribution to the EDM, $d_\ell^Z$, is
\begin{align}
d_\ell^Z(\mu) = e\bar \theta \times \frac{m_\ell}{\pi^2 f_\pi^2} \frac{\alpha}{4\pi} \frac{G_F f_\pi^2}{\sqrt{2}}
	\left[
		c_\pi^{(b)} \log \frac{\Lambda}{\mu}
		+ c_\pi^{(K)} \log \frac{m_K}{\mu}
	\right], \label{eq:EW contribution} \\
%
\rightarrow \quad d_e^Z \simeq -(1-3) \times 10^{-30} \bar\theta \, \text{e-cm}\, ,\qquad
d_\mu^Z \simeq -(2-7) \times 10^{-28} \bar\theta \, \text{e-cm} \, .
\end{align}
For the numerical evaluation, we took $\Lambda = 4\pi f_\pi$ and varied $\mu$ from $m_\pi/2$ to $m_\pi$.
Thus, the electroweak contribution is smaller than the hadronic light-by-light diagrams by two orders of magnitude,
and we can safely neglect this contribution.

\subsubsection{$\boldsymbol\eta^0$ contribution: baryon loop}

Performing a similar calculation as for the $\pi^0$ contribution, the baryon loop contribution in $\eta^0$ 
exchange diagrams is estimated to be
\begin{align}
d_\ell^{\eta,b}(\mu)
= e \bar\theta \times \frac{12 m_\ell}{\pi^2 f_\pi^2}\left( \frac{\alpha}{4\pi} \right)^3
		c_\eta^{(b)} \log^2 \frac{\Lambda}{\mu} \, .
\end{align}
To evaluate the numerical value, we take $\Lambda = 4\pi f_\pi$ and vary $\mu$ from $m_\eta/2$ to $m_\eta$.
This gives,
\begin{align}
d_e^{\eta,b} \simeq -(0.8-3.1) \times 10^{-28} \bar\theta \, \text{e-cm} \, , \qquad
d_\mu^{\eta,b} \simeq -(0.2-0.6) \times 10^{-25} \bar\theta \, \text{e-cm} \, . \label{eq:eta contribution(numerical)}
\end{align}
This contribution is smaller but comparable to $d_\ell^\pi$. 

\subsubsection{$\boldsymbol\eta^0$ contributions: $\boldsymbol\pi$ and $\boldsymbol K$ loop}
\label{eq:eta pi pi contribution}

Here we discuss $\eta^0$ exchange diagram with $\pi^\pm$ or  $K^\pm$ loop.
Since $\eta^0$ is the heaviest particle in the diagram, we integrate out $\eta^0$ first.
After integrating $\eta^0$ out, we obtain the following effective interaction,
\begin{align}
{\cal L}_{\rm eff.}
\ni
 \frac{1}{f_\pi^2} C_3 \pi^+ \pi^- F_{\mu\nu} \tilde F^{\mu\nu},\qquad
C_3 = -  \frac{\bar\theta m_*}{\bar m} \frac{\alpha}{12\pi} \frac{m_\pi^2}{m_\eta^2} \, .
\end{align}
This operator can generate lepton EDM at three-loop level. Its contribution has the same power 
of the coupling constant and loop factor as in Eq.~(\ref{eq:pion contribution}).
However, the operator mixing between $C_3$ and $d_\ell$ gives a contribution which is proportional to $C_3 \log m_\eta/m_\pi$,
and this is not the leading contribution in the sense of chiral logarithms.
So we do not consider this contribution here. We do not discuss $K^\pm$ loop effect for the same 
reason.

\section{Summary}\label{sec:conclusion}

It can be seen from the calculation above that the hadronic light-by-light diagrams give
dominant contribution to the Wilson Coefficients of the lepton EDM operators.
Therefore,
by using Eqs.~(\ref{eq:pion contribution(numerical)}, \ref{eq:eta contribution(numerical)}),
 we obtain
\begin{align}
d_e \simeq -(2.2-8.6) \times 10^{-28} \bar\theta \,  \text{e-cm} \, , \qquad
d_\mu \simeq -(0.5-1.8) \times 10^{-25} \bar\theta \,  \text{e-cm} \, .
\end{align}
Considering the upper bound on neutron EDM, $|d_n| < 3.0 \times 10^{-26} \, \text{e-cm}$ \cite{Afach:2015sja}, and 
the resulting upper bound on the size of $\bar\theta$ to be $|\bar\theta| \lesssim 10^{-10}$ \cite{Vicari:2008jw, Dubbers:2011ns}, 
the $\bar\theta$ contribution to the lepton EDMs are bounded as
\begin{align}
|d_e| < 8.6 \times 10^{-38} \,  \text{e-cm} \, , \qquad
|d_\mu| < 1.8 \times 10^{-35} \,  \text{e-cm}\, .
\label{last-eq}
\end{align}
As mentioned before,  $d_\mu$ in Eq.~(\ref{last-eq}) above is actually equal to the muon EDM which is measured by the spin 
precession of muon. The experiments in J-PARC \cite{Iinuma:2011zz} and Fermilab \cite{Chislett:2016jau} will have 
sensitivities to measure muon EDM up to about $d_\mu = 10^{-20}$ and $10^{-21} \,  \text{e-cm}$ respectively.
Therefore, $\bar\theta$ contribution to the muon EDM is much below sensitivities of the near future experiments.

On the other hand, the bound on $d_e$ from the ACME experiment \cite{Baron:2013eja} assumes vanishing Wilson 
Coefficient for the operator $(\bar e i\gamma^5 e)(\bar N N)$, and hence, their bound should not be compared directly 
with the Wilson Coefficient $d_e$ of the electron EDM operator given in Eq.~(\ref{edm-def}).

As the CKM contribution to $d_e \, (d_\mu)$ is estimated to be of the order of $10^{-44} \, (10^{-42})$ e-cm \cite{Pospelov:2013sca}, 
our calculation shows that the $\bar\theta$ contribution to lepton EDMs,  even after taking into account the strong bound 
on $\bar\theta$ from neutron EDM, can be larger than the CKM contribution by many orders of magnitude.

\section*{Acknowledgements}
We thank Rick Sandeepan Gupta, Gilad Perez and Masahiro Takimoto for discussions. 
We are especially thankful to Maxim Pospelov for discussions which led to the correction of 
an important mistake in the first arXiv version of this paper.


\bibliographystyle{utphys}
\bibliography{ref}
\end{document}